%% file: concrete_2013-12-03.tex
\documentclass[10pt]{article}

\usepackage{amssymb,amsbsy}
\usepackage{amsmath}
\usepackage{amsthm}
\usepackage{graphics,graphicx}
\usepackage[T1]{fontenc}
\usepackage{color}
\usepackage{booktabs}
\usepackage{array}
\usepackage{lineno}

\newcolumntype{L}[1]{>{\raggedright\let\newline\\\arraybackslash\hspace{0pt}}m{#1}}
\newcolumntype{C}[1]{>{\centering\let\newline\\\arraybackslash\hspace{0pt}}m{#1}}
\newcolumntype{R}[1]{>{\raggedleft\let\newline\\\arraybackslash\hspace{0pt}}m{#1}}

\input{definitions}

\title{A simple and robust elastoplastic \\ constitutive model for concrete}

\author{F. Poltronieri$^1$, A. Piccolroaz$^1$, D. Bigoni$^1$\footnote{
Corresponding author. Email: bigoni@ing.unitn.it; webpage http://www.ing.unitn.it/~bigoni/; Fax 00390461282599; Tel. 00390461282507
}, S. Romero Baivier$^2$
\\
\\
{\it $^1$ Department of Civil, Environmental and Mechanical Engineering}
\\ {\it University of Trento, Via Mesiano 77, I-38123 Trento, Italia}
\\ {\it $^2$ Vesuvius Group S.A., Rue de Douvrain 17; 7011 Ghlin, Belgium}
}

\date{}

\begin{document}

\maketitle

\begin{abstract}
\noindent
An elasto-plastic model for concrete, based on a recently-proposed yield surface and simple hardening laws, is formulated, implemented, 
numerically tested and validated against available test results. The yield surface is smooth and particularly suited to represent the behaviour of rock-like 
materials, such as concrete, mortar, ceramic and rock. 
A new class of isotropic hardening laws is proposed, which can be given both an incremental and the corresponding finite form. These laws describe a smooth transition from 
linear elastic to plastic behaviour, incorporating linear and nonlinear hardening, and may approach the perfectly plastic limit in the latter case.
The reliability of the model is demonstrated by its capability of correctly describing the results yielded by a number of well documented triaxial tests on concrete subjected to various 
confinement levels. Thanks to its simplicity, the model turns out to be very robust and well suited to be used in complex design situations, as those involving dynamic loads.
\end{abstract}

\vspace{5 mm}


\section{Introduction}

The mechanical behaviour of concrete is rather complex -- even under monotonic and quasi-static loading -- because of a number of factors:
(i.) highly {\it nonlinear}  and (ii.) strongly 
{\it inelastic} response, (iii.) {\it anisotropic} and (iv.) eventually {\it localized} damage accumulation with (v.) {\it stiffness degradation}, (vi.) {\it 
contractive} and subsequently {\it dilatant} volumetric strain, leading to (vii.) {\it progressively severe cracking}. 
This complexity is the macroscale counterpart 
of several concurrent and cooperative or antagonist micromechanisms of damage and stiffening, as for instance, pore collapse, microfracture opening and extension or 
closure, aggregate debonding, and interfacial friction. 
It can therefore be understood that the constitutive modelling of concrete (but also of similar materials such as rock, soil and ceramic) has been the subject of an 
impressive research effort, which, broadly speaking, falls within the realm of elastoplasticity\footnote{
The research on elastoplastic modelling of concrete 
has reached its apex in the seventies and eighties of the past century, when so many models have been proposed that 
are now hard to even only be summarized (see among others, Bazant et al., 2000; Chen and Han, 1988; Krajcinovic et al., 1991;
Yazdani and Schreyer, 1990). 
Although nowadays other approaches are preferred, like those based on
particle mechanics (Schauffert and Cusatis, 2012), the aim of the present article is to formulate a relatively simple and robust model, something that is difficult to be achieved
with advanced models.
}, 
where the term \lq plastic' is meant to include the damage as a specific inelastic mechanism.
In fact, elastoplasticity is a theoretical framework allowing the possibility of a phenomenological description of all the above-mentioned constitutive features of concrete in terms of: (i.) 
yield function features, (ii.) coupling elastic and plastic deformation, (iii.) flow-rule nonassociativity, and (iv.) hardening sources and rules (Bigoni, 2012).
However, the usual problem arising from a refined constitutive description in terms of
elastoplasticity is the complexity of the resulting model, which may 
lead to several numerical difficulties related to the possible presence of yield surface corners, discontinuity of hardening, lack of self-adjointness due to 
nonassociativity and failure of ellipticity of the rate equations due to strain softening. As a consequence, refined models often lack numerical robustness or slow 
down the numerical 
integration to a level that the model becomes of awkward, if not impossible, use. 
A \lq minimal' and robust constitutive model, not obsessively accurate but able to capture the essential phenomena related to the progressive damage occurring 
during monotonic loading of concrete, is a necessity to treat complex load situations. 

The essential \lq ingredients' of a constitutive model are a convex, 
smooth yield surface capable of an excellent interpolation of data and a hardening law describing a smooth transition between elasticity and a perfectly plastic 
behaviour. Exclusion of strain softening, together with flow rule associativity, is the key to preserve ellipticity, and thus well-posedness of the problem. 
In the present paper, an elastoplastic model is formulated, based on the so-called \lq BP yield surface' (Bigoni and Piccolroaz, 2004; Piccolroaz and Bigoni, 2009; Bigoni, 2012) which is shown
to correctly describe the damage envelope of concrete, and on an infinite class of isotropic hardening rules (given both in incremental form and in the corresponding 
finite forms), depending on a hardening parameter. 
Within a certain interval for this parameter hardening is unbounded (with linear hardening obtained as a limiting case), while outside this range a smooth 
hardening/perfectly-plastic transition is described. 
The proposed model does not describe certain phenomena which are known to occur in concrete, such as for instance anisotropy of damage, softening, and elastic 
degradation, but provides a simple and robust tool, which is shown to correctly represent triaxial test results at high confining pressure.

\section{Elastoplasticity and the constitutive model}

This section provides the elastoplastic constitutive model in terms of incremental equations. The form of the yield surface is given, depending on the stress invariants and a set of material parameters. A new class of hardening laws is formulated in order to describe a smooth transition from linear elastic to plastic 
behaviour.

\subsection{Incremental constitutive equations}

The decomposition of the strain into the elastic ($\bepsilon_e$) and plastic ($\bepsilon_p$) parts as
\beq
\bepsilon = \bepsilon_e + \bepsilon_p
\eeq
yields the incremental elastic strain in the form
\beq
\dot{\bepsilon}_e = \dot{\bepsilon} - \dot{\bepsilon}_p.
\label{dotepsilon}
\eeq
The `accumulated plastic strain' is defined as follows
\beq
\pi_a = \int^{t}_{0} \left|\dot{\bepsilon}_p\right| d\tau,
\eeq
where $t$ is the time-like variable governing the loading increments. 
Introducing the flow rule
\beq
\dot{\bepsilon}_p=\dot{\lambda}\bP,
\label{flow}
\eeq
where $\bP$ is the gradient of the plastic potential, we obtain that the rate of the accumulated plastic strain is proportional to the plastic multiplier ($\dot{\lambda}\geq0$) as 
\beq
\dot{\pi}_a = \dot{\lambda}\left|\bP\right|.
\eeq
A substitution of eq. \eqref{dotepsilon} and eq. \eqref{flow} into the incremental elastic constitutive equation relating the increment of stress $\dot{\bsigma}$ to the increment of elastic strain $\dot{\bepsilon}_e$ through a fourth-order elastic tensor $\fE$ as $\dot{\bsigma} = \fE \dot{\bepsilon}_e,$ 
yields
\beq
\dot{\bsigma} = \fE\dot{\bepsilon}-\dot{\lambda}\fE\bP.
\label{sigmapunto}
\eeq
Note that, for simplicity, reference is made to isotropic elasticity, so that the elastic tensor $\fE$ is defined in terms of elastic modulus $E$ and Poisson's ratio $\nu$. 

During plastic loading, the stress point must satisfy the yield condition $F(\bsigma,\bk) = 0$ at every time increment, so that the Prager consistency can be written as
\beq
\dot{F} = \bQ \cdot \dot{\bsigma} + \frac{\partial{F}}{\partial\bk} \cdot \dot{\bk} = 0,
\label{Prager}
\eeq
where $\bQ = \partial{F}/\partial\bsigma$ is the yield function gradient and $\bk$ is the hardening parameters vector.

By defining the hardening modulus $H(\bsigma,\bk)$ as
\beq
\frac{\partial{F}}{\partial\bk}\cdot\dot{\bk} = -\dot{\lambda}H(\bsigma,\bk),
\eeq
eq. \eqref{Prager} can be rewritten in the form 
\beq
\bQ\cdot\dot{\bsigma} - \dot{\lambda}H = 0
\label{Prager2}
\eeq
Further, the plastic multiplier can be obtained from eq. \eqref{sigmapunto} and eq. \eqref{Prager2}
\beq
\dot{\lambda}=\frac{\bQ\cdot\fE\dot{\bepsilon}}{H+\bQ\cdot\fE\bP}.
\label{lambdapunto}
\eeq
Finally, a substitution of eq. \eqref{lambdapunto} into eq. \eqref{sigmapunto} yields the elasto-plastic constitutive equations in the rate form
\beq
\dot{\bsigma}=\fE\dot{\bepsilon}-\frac{\bQ\cdot\fE\dot{\bepsilon}}{H+\bQ\cdot\fE\bP}\fE\bP,
\eeq
where, for simplicity, the associative flow rule  $\bP = \bQ$, will be adopted in the sequel.

\subsection{The BP yield surface}

The following stress invariants are used in the definition of the BP yield function (Bigoni and Piccolroaz, 2004)
\begin{equation}
p = -\frac{\tr \bsigma}{\ds3}, \quad q = \sqrt{3J_2}, \quad \theta = \frac{1}{3}\cos^{-1} \left(\frac{3\sqrt{3}}{2}\frac{J_3}{J_2^{3/2}}\right),
\end{equation}
where $\theta \in [0, \pi/3]$ is the Lode's angle and
\begin{equation}
J_2 = \frac{1}{2} \tr \bS^2, \quad J_3 = \frac{1}{3} \tr \bS^3, \quad \bS = \bsigma - \frac{\tr\bsigma}{3}\bI,
\end{equation}
in which $\bS$ is the deviatoric stress and $\bI$ is the identity tensor.

The seven-parameters BP yield function $F$ (Bigoni and Piccolroaz, 2004) is defined as
\beq
F(\bsigma) = f(p) + \frac{q}{g(\theta)},
\eeq
in which the pressure-sensitivity is described through the \lq meridian function'
\begin{equation}
f(p) =
-M p_c \sqrt{\left(\frac{p + c}{p_c + c} - \left(\frac{p + c}{p_c + c}\right)^m\right) \left[ 2(1-\alpha)\frac{p + c}{p_c + c} + \alpha \right]}, ~~\text{ if }~~ \frac{p + c}{p_c + c} \in [0,1],
\end{equation}
and $f(p) = + \infty$, if $(p + c)/(p_c + c) \, \notin [0,1]$. 
The Lode-dependence of yielding is described by the \lq deviatoric function' (proposed by Podg\'{o}rski, 1985  and independently by Bigoni and Piccolroaz, 2004)
\beq
1/g(\theta) = \cos \left[\beta\frac{\pi}{6} - \frac{1}{3} \arccos(\gamma \cos 3\theta)\right].
\label{gditeta}
\eeq
The seven, non-negative material parameters
$$
M > 0, \quad p_c > 0, \quad c \geq 0, \quad 0< \alpha < 2, \quad m > 1, \quad 0 \leq \beta \leq 2, \quad 0 \leq \gamma < 1
$$
define the shape of the associated yield surface. In particular, $M$ controls the pressure-sensitivity, $p_c$ and $c$ are the yield strengths under ideal isotropic compression and tension, respectively. 
Parameters $\alpha$ and $m$ define the distortion of the meridian section, while $\beta$ and $\gamma$ model the shape of the deviatoric section.

\subsection{An infinite class of hardening laws}

In order to simulate the nonlinear hardening of concrete, the following class of hardening rules in incremental form is proposed:
\beq
\dot{p}_c = \frac{k_1}{(1 + \delta\pi_a)^n} \dot{\pi}_a,
\label{pcpunto}
\eeq
\beq
\dot{c} = \Omega\dot{p_c},
\label{cpunto}
\eeq
where four material parameters have been introduced: $k_1 > 0$, $\delta \ge 0$, $n > 0$, and $0 < \Omega < 1$.

A substitution of the flow rule \eqref{flow} into \eqref{pcpunto} and \eqref{cpunto} yields
\beq
\label{inc1}
\dot{p}_c = \dot{\lambda}\frac{k_1}{(1 + \delta\pi_a)^n} \left|\bP\right|,
\eeq
\beq
\label{inc2}
\dot{c} = \dot{\lambda}\frac{\Omega k_1}{(1 + \delta\pi_a)^n} \left|\bP\right|.
\eeq
Eqs. (\ref{pcpunto}) and (\ref{cpunto}) can be integrated in order to obtain the hardening laws in finite form as follows
\beq
\label{finita1}
p_c = 
\left\{
\barr{ll}
\ds p_{c0} + \frac{k_1}{\delta} \log(1 + \delta\pi_a), & n = 1, \\[3mm]
\ds p_{c0} + \frac{k_1}{\delta}\frac{(1 + \delta \pi_a)^{n - 1} - 1}{(n - 1) (1 + \delta \pi_a)^{n - 1}}, & n \neq 1.
\earr
\right.
\eeq
\beq
\label{finita2}
c - c_0 = \Omega (p_c - p_{c0}).
\eeq
In order to gain insight on the physical meaning of the proposed class of hardening rules, let us consider the asymptotic behaviour of the finite form 
(\ref{finita1}) as $\pi_a \to 0$, which turns out to be independent of the parameters $\delta$ and $n$ and is given by the formula
\beq
p_c = p_{c0} + k_1 \pi_a + O(\pi_a^2).
\eeq
This formula shows that, at an early stage of plastic deformation, the hardening is linear with a slope governed by the parameter $k_1$.

\begin{figure}[!htb]
\centering
\includegraphics[width=10cm]{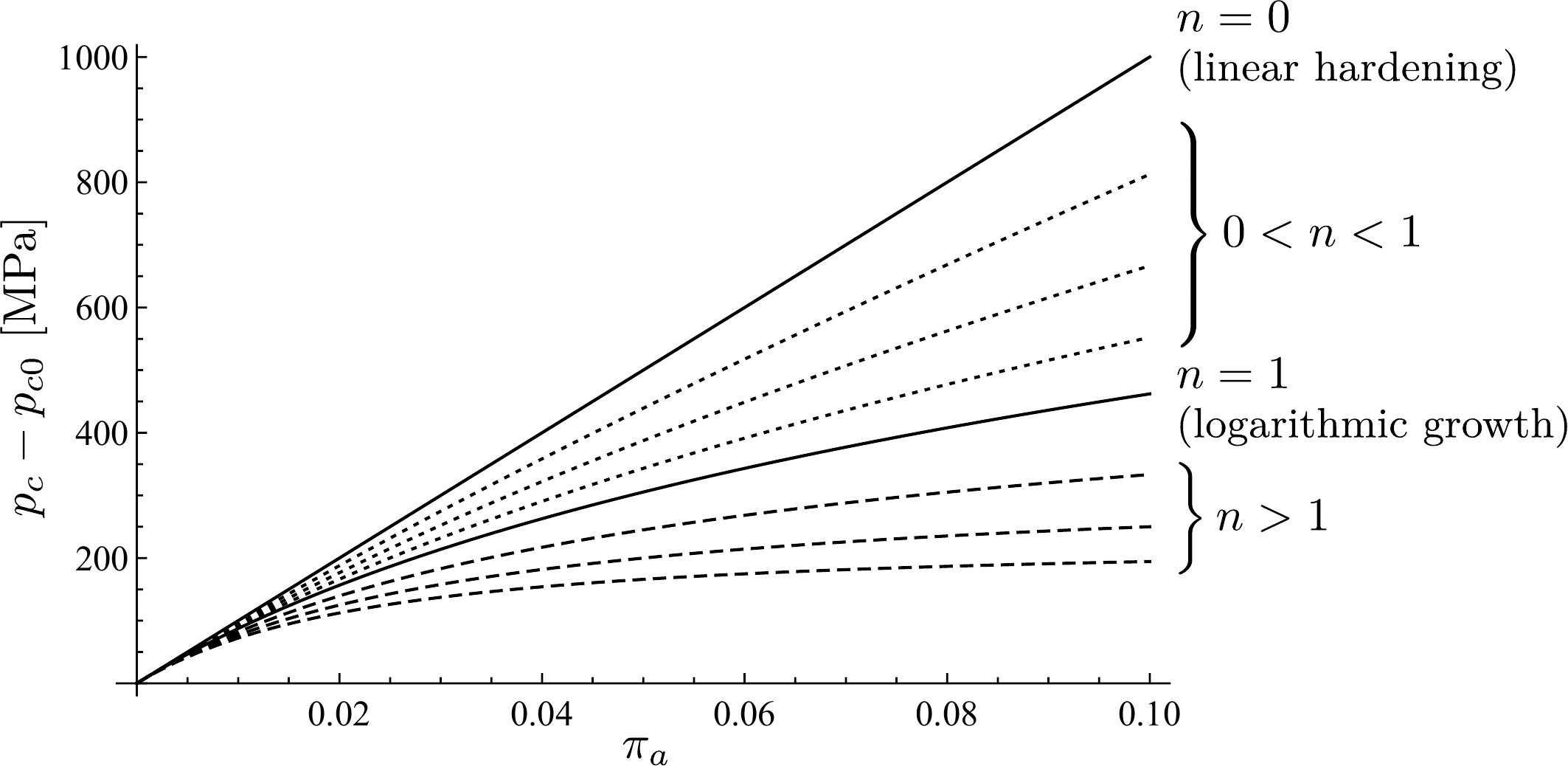}
\caption{\footnotesize{The infinite class of hardening laws corresponding to eq. (\ref{finita1}). Curves are obtained by setting the parameters $k_1$ and $\delta$ to the 
conventional values $k_1 = 10000$ MPa, $\delta = 30$, and varying the parameter $n$.}}
\label{fig:fig00j}
\end{figure}

On the other hand, the hardening at a later stage of plastic deformation is governed by the parameters $\delta$ and $n$. This is made clear by considering the 
asymptotic behaviour of the finite form (\ref{finita1}) as $\pi_a \to \infty$:
\beq
p_c = 
\left\{
\barr{ll}
\ds \frac{k_1}{\delta} \log(\delta \pi_a) + p_{c0} + \frac{k1}{\delta^2 \pi_a} + O(\pi_a^{-2}), & n = 1, \\[3mm]
\ds p_{c0} + \frac{k_1}{(n - 1)\delta} - \frac{k_1}{(n - 1)\delta (\delta \pi_a)^{n - 1}} + O(\pi_a^{-n}), & n \neq 1.
\earr
\right.
\eeq
Note that, for $0< n \leq 1$, $p_c$ turns out to be unbounded and for $n=1$ there is a logarithmic growth of hardening. 
For $n > 1$, $p_c$ is bounded (Fig. \ref{fig:fig00j}) with a limit value given by 
$$
p_{c0} + \frac{k1}{(n - 1)\delta}.
$$

With the definition of the hardening law, the proposed elastoplastic constitutive model is ready to be calibrated, implemented and tested, which is the subject of the following Sections, where two members of the class of hardening laws will be analyzed, namely, $n=1$ and $n=2$, the former corresponding to the logarithmic growth of hardening and the latter approaching the perfectly plastic behaviour.

\section{Calibration of material parameters and comparison with experimental results}

\subsection{Calibration of the BP yield surface}

\begin{figure}[!htb]
\centering
\includegraphics[width=8cm]{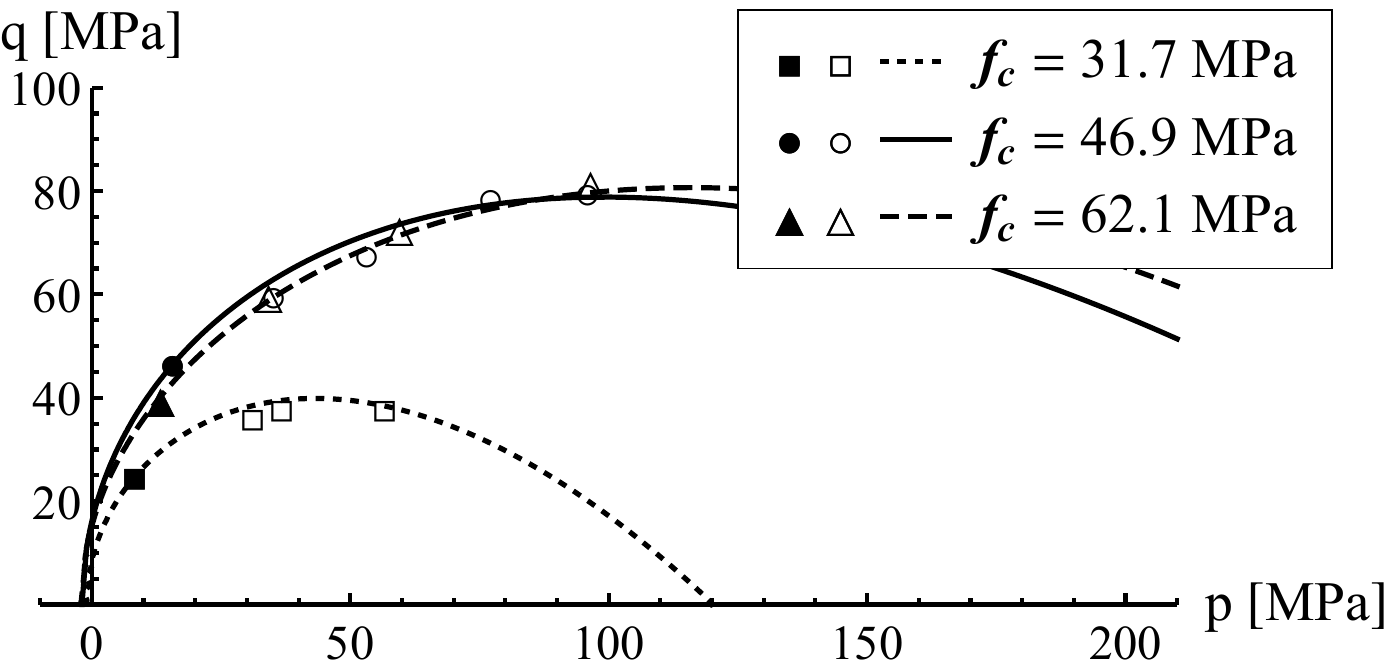}
\caption{\footnotesize{The meridian section ($p$--$q$ plane) of the BP yield surface interpolating experimental results for three concretes (data taken from Kotsovos and  Newman, 1980). Black (white) spots denote results relative to unconfined (confined) compression}}
\label{fig:graf01c}
\end{figure}

\begin{figure}[!htb]
\centering
\includegraphics[width=7cm]{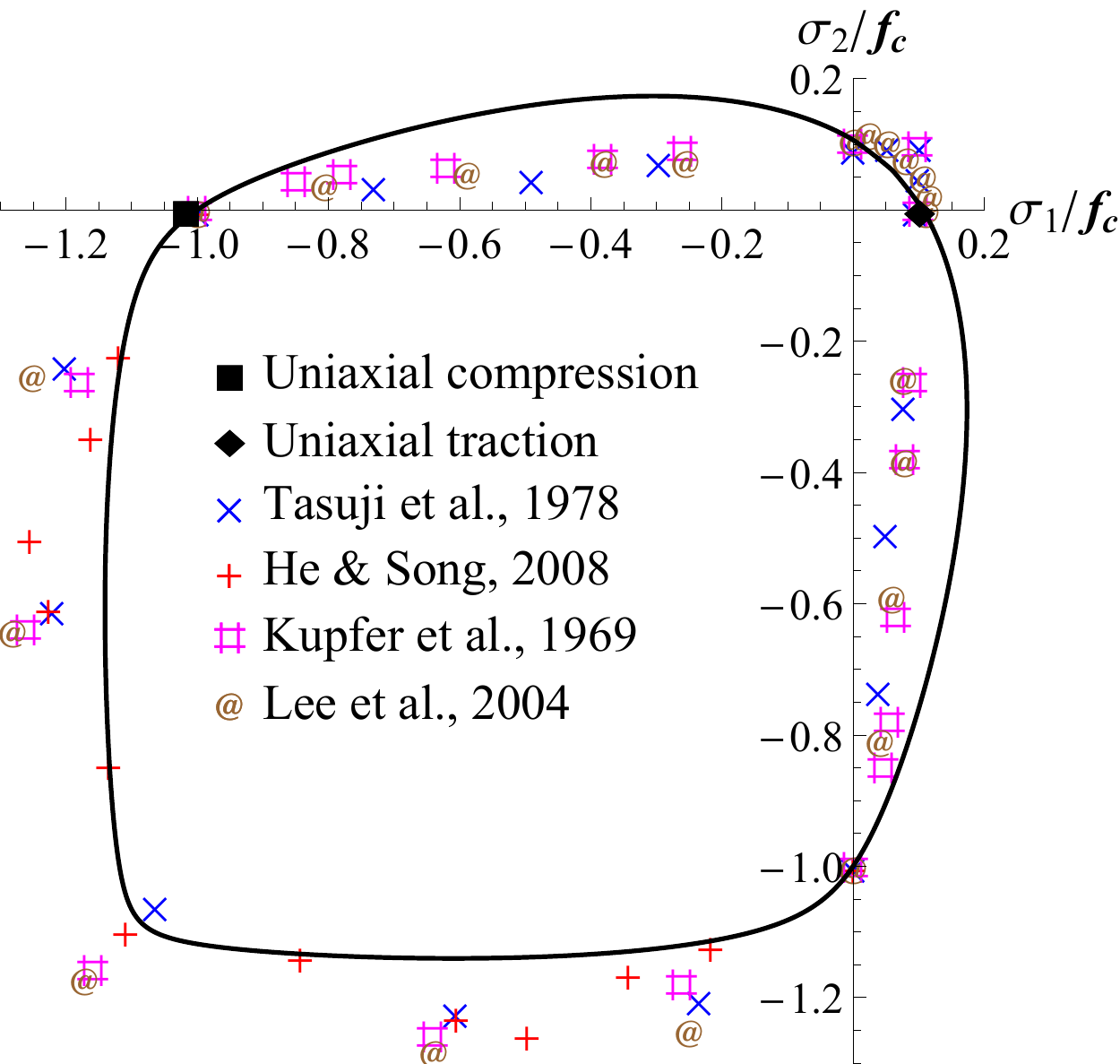}
\caption{\footnotesize{The BP yield surface in the biaxial plane interpolating experimental results for concrete}}
\label{fig:graf02}
\end{figure}

The BP yield surface has been calibrated on the basis of available experiments (He and Song, 2008; Kotsovos and Newman, 1980; Kupfer et al., 1969; Lee et al., 2004; Tasuji et al., 1978), to model the behaviour of concrete in the $p$--$q$ plane (Fig. \ref{fig:graf01c}) and in the biaxial $\sigma_1/f_c$--$\sigma_2/f_c$ plane (Fig. \ref{fig:graf02}), where $f_c$ (taken positive) defines the failure for compressive uniaxial stress. We have assumed a ratio between $f_c$ and the failure stress in uniaxial tension $f_t$ equal to 10. 
%
%

\noindent
The parameter values which have been found to provide the better interpolation of experimental data for the  three concretes (identified by their uniaxial 
compressive strengths $f_c = 31.7$ MPa, $f_c = 46.9 $ MPa, and $f_c = 62.1$ MPa) reported by Kotsovos and Newman (1980) are given in Tab. \ref{tab01}.
%
\begin{table}[!htcb]
\begin{center}
\begin{tabular}{L{25mm} C{12mm} C{12mm} C{12mm} C{12mm} C{12mm} C{12mm} C{12mm}}
\toprule
Concrete's            & $M$      & $m$      & $\alpha$ & $\beta$  & $\gamma$ & $p_c$    & $c$      \\
compression strength  &          &          &          &          &          & (MPa)    & (MPa)    \\
\midrule
$f_c = 31.7$ MPa      & 0.435    & 2.5      & 1.99     & 0.6      & 0.98     & 120      & 1.02     \\
$f_c = 46.9$ MPa      & 0.38     & 2        & 1.99     & 0.6      & 0.98     & 300      & 1.9      \\
$f_c = 62.1$ MPa      & 0.33     & 2.5      & 1.99     & 0.6      & 0.98     & 320      & 2.1      \\
\bottomrule
\end{tabular}
\end{center}
\caption{Yield function parameters employed for the simulation of triaxial compression tests.}
\label{tab01}
\end{table}

\subsection{A validation of the model against triaxial tests}

The above-presented constitutive model has been implemented in a UMAT routine of ABAQUS Standard Ver. 6.11-1 (Simulia, Providence, RI). 

Triaxial compression tests at high pressure on cylindrical concrete specimen (experimental data taken from Kotsovos and Newman, 1980) have been simulated 
using two different hardening laws of the class (\ref{pcpunto})--(\ref{cpunto}), namely $n = 1$, corresponding to logarithmic hardening, and $n = 2$, which 
provides a smooth transition to perfect plasticity.
The values of parameters (identifying  elastic behaviour and hardening) used in the simulations are reported in Tab. \ref{tab02}.
%
\begin{table}[!htcb]
\begin{center}
\begin{tabular}{L{25mm} C{15mm} C{15mm} C{15mm} C{15mm} C{15mm}}
\toprule
Material              & $E$            & $\nu$          & $k_1$            & $\delta$       & $\Omega$       \\
                      & (MPa)          &                & (MPa)            &                &                \\
\midrule
$f_c = 31.7$ MPa      &                &                &                  &                &                \\
\hspace{13mm} $n = 1$ &  13000         &  0.18          &  25000           &  220           &  0.0085        \\
\hspace{13mm} $n = 2$ &  13000         &  0.18          &  25000           &  80            &  0.0085        \\
\midrule
$f_c = 46.9$ MPa      &                &                &                  &                &                \\
\hspace{13mm} $n = 1$ &  11200         &  0.18          &  223000          &  2300          &  0.00633       \\
\hspace{13mm} $n = 2$ &  11200         &  0.18          &  100000          &  190           &  0.00633       \\
\midrule
$f_c = 62.1$ MPa      &                &                &                  &                &                \\
\hspace{13mm} $n = 1$ &  43000         &  0.18          &  210000          &  220           &  0.00656       \\
\hspace{13mm} $n = 2$ &  43000         &  0.18          &  210000          &  220           &  0.00656       \\
\bottomrule
\end{tabular}
\end{center}
\caption{Elastic and hardening parameters employed for the simulation of triaxial compression tests.}
\label{tab02}
\end{table}

\begin{figure}[!htcb]
\centering
\includegraphics[width=11cm]{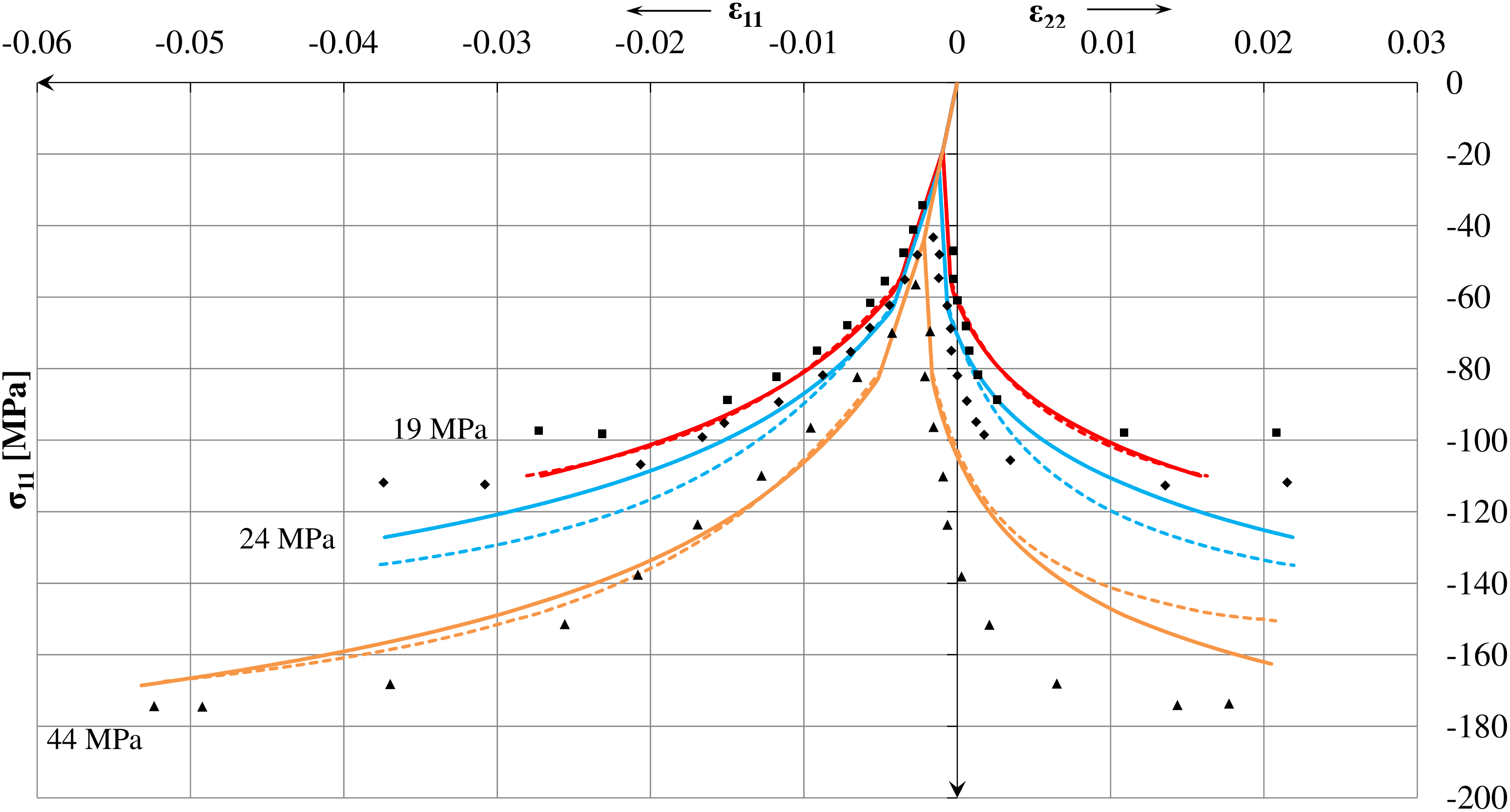}
\caption{\footnotesize{Comparison between the simulation of triaxial tests at different confining pressures and the relative experiments (taken from Kotsovos and  Newman, 1980, concrete with $f_c = 31.7$ MPa). The solid [the dashed] lines correspond to the case $n = 1$ [$n = 2$].}}
\label{fig:provetriax2}
\end{figure}

\begin{figure}[!htcb]
\centering
\includegraphics[width=11cm]{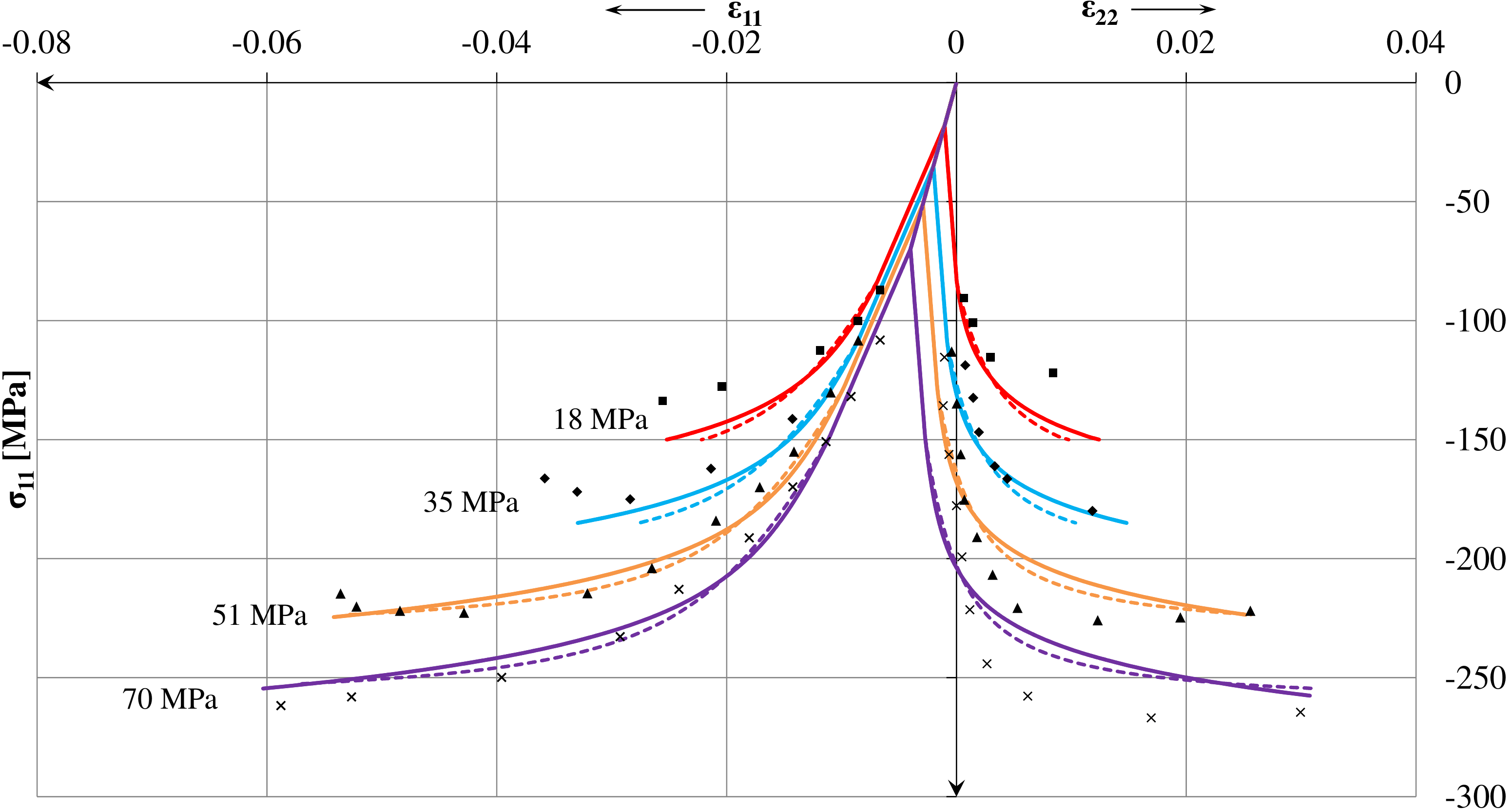}
\caption{\footnotesize{Comparison between the simulation of triaxial tests at different confining pressures and the relative experiments (taken from Kotsovos and  Newman, 1980, concrete with $f_c = 46.9$ MPa). The solid [the dashed] lines correspond to the case $n = 1$ [$n = 2$].}}
\label{fig:provetriax}
\end{figure}

\begin{figure}[!htcb]
\centering
\includegraphics[width=11cm]{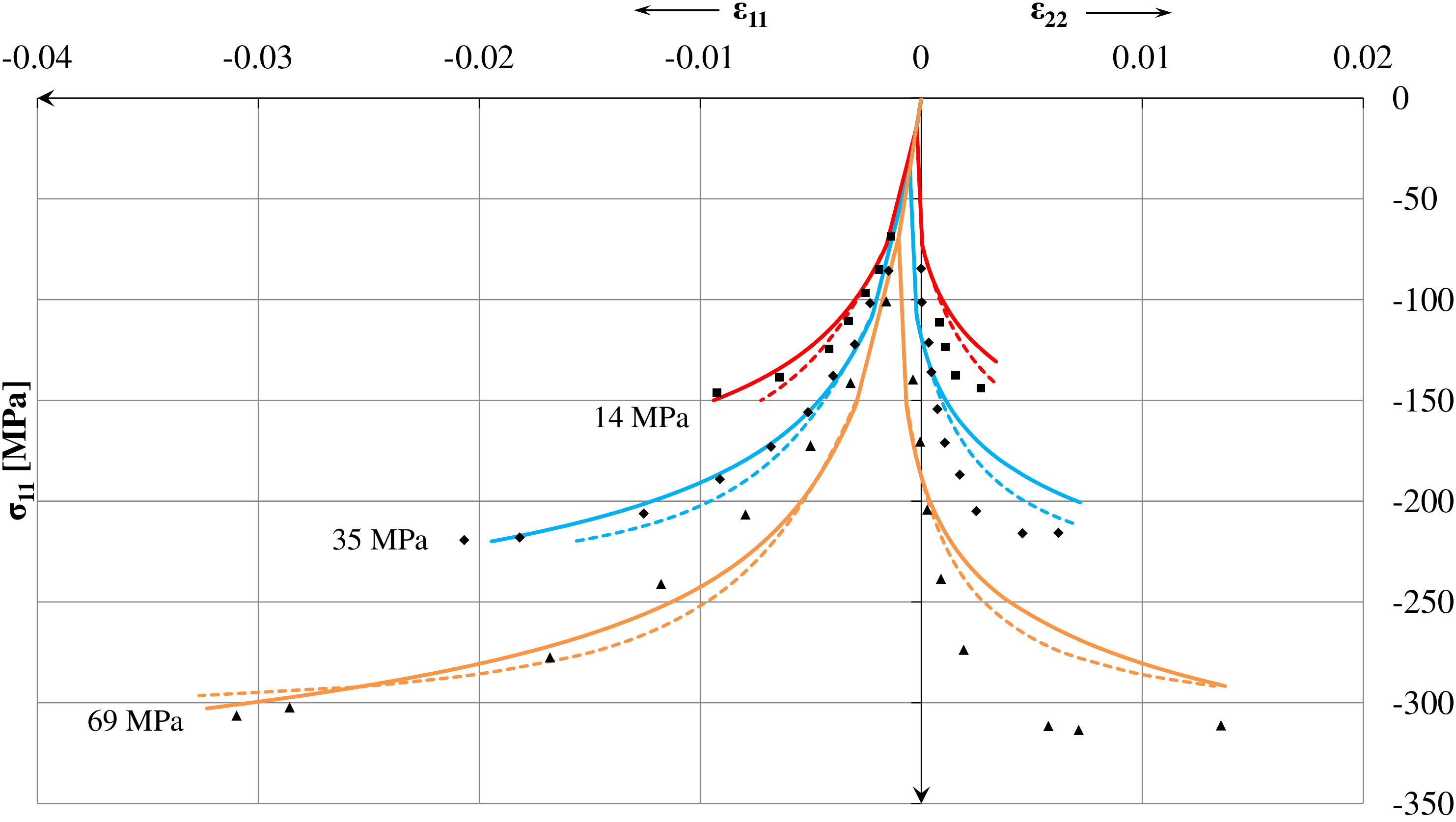}
\caption{\footnotesize{Comparison between the simulation of triaxial tests at different confining pressures and the relative experiments (taken from Kotsovos and  Newman, 1980, concrete with $f_c = 62.1$ MPa). The solid [the dashed] lines correspond to the case $n = 1$ [$n = 2$].}}
\label{fig:provetriax3}
\end{figure}

Results of the numerical simulations (one axisymmetric biquadratic finite element with eight nodes --so called \lq CAX8 '-- has been employed) are presented in 
Figs. \ref{fig:provetriax2}--\ref{fig:provetriax3},  reporting the axial stress $\sigma_{11}$ plotted in terms of axial $\epsilon_{11}$ and radial $\epsilon_{22}$ 
strains. The three figures refer to the three different concretes, identified by their different uniaxial compressive strength. Different colours of the lines 
denote simulations performed at different confining pressure, namely: $\{$19, 24, 44$\}$ MPa for Fig.\ref{fig:provetriax2}, $\{$18, 35, 51, 70$\}$ MPa for Fig. 
\ref{fig:provetriax}, and $\{$14, 35, 69$\}$ MPa for Fig. \ref{fig:provetriax3}.

The solid lines correspond to the logarithmic hardening laws (\ref{finita1}) and (\ref{finita2}) with $n = 1$. The dashed lines correspond to the hardening 
laws (\ref{finita1}) and (\ref{finita2}) with $n = 2$.
%
%
%

It is clear from the figures that both the hardening laws perform correctly, but the hardening rule with $n=1$ leads to a more stable 
numerical behaviour, while with $n=2$ failure can be approached.

\newpage

\section{Conclusions}

A reasonably simple inelastic model for concrete has been formulated on the basis of a recently-proposed yield function and a class of hardening rules, which may describe a smooth transition from linear elastic to perfectly plastic behaviour. The model has been implemented into a numerical routine and successfully validated against available triaxial experiments at different confining pressures. 
Although the presented constitutive approach cannot compete in accuracy with more sophisticated models, it has been proven to be robust for numerical calculations, so that it
results particularly suitable for complex design situations where easy of calibration, fast convergence and stability 
become important factors.

\vspace{10 mm}
\noindent
\section*{Acknowledgments}
Part of this work was prepared during secondment of D.B. at Enginsoft (TN). 
D.B. and A.P. gratefully acknowledge financial support from European Union FP7 project under contract number PIAP-GA-2011-286110-INTERCER2.
S.R.B. gratefully acknowledges financial support from European Union FP7 project under contract number PIAPP-GA-2013-609758-HOTBRICKS.

\end{document}

%% file: definitions.tex
\newcommand{\beq}{\begin{equation}}
\newcommand{\eeq}{\end{equation}}

\newcommand{\beqar}{\begin{eqnarray}}
\newcommand{\eeqar}{\end{eqnarray}}
\newcommand{\bit}{\begin{itemize}}
\newcommand{\eit}{\end{itemize}}
\newcommand{\benum}{\begin{enumerate}}
\newcommand{\eenum}{\end{enumerate}}
\newcommand{\barr}{\begin{array}}
\newcommand{\earr}{\end{array}}
\def\ds{\displaystyle}

\def\XXint#1#2#3{{\setbox0=\hbox{$#1{#2#3}{\int}$}
   \vcenter{\hbox{$#2#3$}}\kern-.5\wd0}}

\def\b0{\mbox{\boldmath $0$}}

\def\bk{\mbox{\boldmath $k$}}

\def\bI{\mbox{\boldmath $I$}}

\def\bP{\mbox{\boldmath $P$}}
\def\bQ{\mbox{\boldmath $Q$}}

\def\bS{\mbox{\boldmath $S$}}

\newcommand{\bsigma}{\mbox{\boldmath $\sigma$}}

\newcommand{\bepsilon}{\mbox{\boldmath $\epsilon$}}

\def\f0{\ensuremath{\mathbb{O}}}

\def\fE{\ensuremath{\mathbb{E}}}

\newcommand{\tr}{\mathop{\mathrm{tr}}}

\def\ACIJ{{\it ACI J.}\ }

\def\IJSS{{\it Int.\ J.\ Solids Struct.}\ }

%% file: concrete_2013-12-03.bbl
\begin{thebibliography}{}

\bibitem{abaqus}
Abaqus 6.10, 2010. User Manual Available from http://simulia.com

\bibitem{bazant} Bazant, Z.P., Adley, M.D., Carol, I., Jirasek, M., Akers, S.A., Rohani, B., Cargile, J.D., Caner, F.C. (2000) Large-strain generalization of microplane model for concrete and application. J. Eng. Mech-ASCE  126, 971-980.

\bibitem{big}
Bigoni, D. (2012)  
{\it Nonlinear Solid Mechanics. Bifurcation Theory and Material Instability.} 
Cambridge University Press.

\bibitem{bp}
Bigoni, D., Piccolroaz, A. (2004)
Yield criteria for quasibrittle and frictional materials. 
\IJSS {\bf 41}, 2855-2878.

\bibitem{chen}
Chen, W.F., Han, D.J. (1988) 
{\it Plasticity for structural engineers.} 
Springer-Verlag. 

\bibitem{he}
He, Z., Song, Y. (2008)
Failure mode and constitutive model of plain high-strength high-performance concrete under biaxial compression after exposure to high temperatures.
{\it Acta Mech. Solida Sinica} {\bf 21}, 149-159.

\bibitem{kotsovosnewman}
Kotsovos, M. D., Newman, J. B. (1980)
Mathematical description of deformational behavior of concrete under generalized stress beyond ultimate strength.
\ACIJ {\bf 77}, 340-346.

\bibitem{kraj} Krajcinovic, D., Basista, M. and Sumarac, D. (1991) Micromechanically inspired phenomenological damage model. J. Appl. Mech. ASME 58, 305-310.

\bibitem{kupfer}
Kupfer, H., Hilsdorf, H.K., Rusch, H. (1969) 
Behaviour of concrete under biaxial stress. 
{\it Proc. ACI} {\bf 66}, 656-666.

\bibitem{lee}
Lee, S.-K., Song, Y.-C., Han, S.-H. (2004)
Biaxial behavior of plain concrete of nuclear containment building.
{\it Nucl. Eng. Des.} {\bf 227}, 143-153.

\bibitem{pic}
Piccolroaz, A., Bigoni, D. (2009) 
Yield criteria for quasibrittle and frictional materials: a generalization to surfaces with corners. 
\IJSS {\bf 46}, 3587-3596.

\bibitem{podo2} 
Podg\'{o}rski, J. (1985) General failure criterion for isotropic media. {\it J. Eng. Mech. ASCE} 111, 188-199.

\bibitem{cusa} Schauffert, E.A. and Cusatis, G. (2012) Lattice Discrete Particle Model for Fiber-Reinforced Concrete. I: Theory. J. Eng. Mech. ASCE  138, 826-833. 


\bibitem{tas}
Tasuji, M.E., Slate, F.O., Nilson, A.H. (1978) 
Stress–strain response and fracture of concrete in biaxial loading. 
\ACIJ {\bf 75}, 306-312.

\bibitem{yaz}
Yazdani, S. and Schreyer, H.L. (1990)  Combined plasticity and damage mechanics model for plain concrete. J. Eng. Mech-ASCE, 116, 1435-1450.

\end{thebibliography}
